\documentclass[aps,prl,twocolumn,showpacs,preprintnumbers,amsmath,amssymb,amsfonts,reprint]{revtex4-1}
\usepackage{graphicx}

\usepackage[english]{babel}
\usepackage{hyperref}
\usepackage{xcolor}
\hypersetup{
    colorlinks,
    linkcolor={red!50!black},
    citecolor={blue!50!black},
    urlcolor={blue!80!black}
}
\usepackage[english]{cleveref}
\usepackage[]{units}
\usepackage[]{todonotes}
\begin{document}

\title{Observation of Laser Power Amplification in a Self-Injecting Laser Wakefield Accelerator}

\author{M.J.V.~Streeter$^{1,2,3}$}
\author{S.~Kneip$^{3}$}
\author{M.S.~Bloom$^{3}$}
\author{R.A.~Bendoyro$^{4}$}
\author{O.~Chekhlov$^{5}$} 
\author{A.E.~Dangor$^{3}$}
\author{A.~D\"opp$^{3,7,8}$}
\author{C.J.~Hooker$^{5}$}
\author{J.~Holloway$^{6}$}
\author{J.~Jiang$^{4}$}
\author{N.C.~Lopes$^{3,4}$}
\author{H.~Nakamura$^{3}$}
\author{P.A.~Norreys$^{5}$}
\author{C.A.J.~Palmer$^{1,2}$}
\author{P.P.~Rajeev$^{5}$}
\author{J.~Schreiber$^{7,8}$}
\author{D.R.~Symes$^{5}$}
\author{A.G.R.~Thomas$^{1,2,9}$}
\author{M.~Wing$^{6}$}
\author{S.P.D.~Mangles$^{3}$}
\author{Z.~Najmudin$^{3,*}$}

\affiliation{$^{1}$ The Cockcroft Institute, Keckwick Lane, Daresbury, WA4 4AD, United Kingdom}
\affiliation{$^{2}$ Physics Department, Lancaster University, Lancaster LA1 4YB, United Kingdom}
\affiliation{$^{3}$John Adams Institute for Accelerator Science, The Blackett Laboratory, Imperial College London, London, SW7 2AZ, United Kingdom}
\affiliation{$^{4}$GoLP/Instituto de Plasmas e Fus\~{a}o Nuclear, Instituto Superior T\'{e}cnico, Universidade de Lisboa, Lisboa 1049-001, Portugal}
\affiliation{$^{5}$Central Laser Facility, Rutherford Appleton
Laboratory, Chilton, Oxon, OX11 0QX, United Kingdom}
\affiliation{$^{6}$High Energy Physics Group, University College London, London WC1E 6BT, United Kingdom}
\affiliation{$^{7}$Fakult\"at f\"ur Physik, Ludwig-Maximilians-Universit\"at M\"unchen, am Coulombwall 1, D-85748 Garching, Germany}
\affiliation{$^{8}$Max-Planck-Institut f\"ur Quantenoptik, Hans-Kopfermann-Str.~1, D-85748 Garching, Germany}
\affiliation{$^{9}$Center for Ultrafast Optical Science, University of Michigan, Ann Arbor, MI 48109-2099, USA}
\date{\today}

\begin{abstract}
We report on the depletion and power amplification of the driving laser pulse in a strongly-driven laser wakefield accelerator. 
Simultaneous measurement of the transmitted pulse energy and temporal shape indicate an increase in peak power from \unit[$187 \pm 11$]{TW} to a maximum of \unit[$318 \pm 12$]{TW} after \unit[13]{mm} of propagation in plasma density of \unit[$0.9 \times 10^{18}$]{cm$^{-3}$}. The power amplification is correlated with the injection and acceleration of electrons in the nonlinear wakefield.  This process is modeled by including localized redshift and subsequent group delay dispersion at the laser pulse front.
\end{abstract}
\pacs{41.75.Jv, 42.65.Jx, 52.38.Dx, 52.38.Hb, 52.38.Kd, 52.65.Rr }

\maketitle


Laser wakefield accelerators (LWFA) \cite{Tajima1979PRL} can now produce electron beams with particle energies greater than GeV from centimeter scale interaction lengths \cite{Leemans2014PRL, Wang2013NC, Kim2013PRL}. 
In an LWFA, a high-intensity laser pulse propagating through a plasma initiates a plasma wave, which exhibits extremely high longitudinal electric fields. 
Numerous methods have been demonstrated to inject particles within a LWFA \cite{Esarey2009RMP,RowlandsRees2008PRL,Pak2010PRL,McGuffey2010PRL}. 
Of these, self-injection in the highly nonlinear `bubble' or `blowout', regime \cite{Mangles2004N,Geddes2004N,Faure2004N,Leemans2014PRL} is amongst the simplest and thus most common.
By using self- \cite{Thomas2007PRL} or external guiding \cite{Spence2000PRE}, it is possible to maintain the LWFA far beyond the normal Rayleigh diffraction length. 
However, the eventual energy gain of electrons by the wakefield is limited either by \emph{dephasing}, or by \emph{depletion} of the driving laser pulse \cite{Lu2007PRSTAB}.

Dephasing occurs when electrons outrun the wakefield, which is usually said to move at the linear group velocity of the laser pulse in the plasma, $v_g = c \sqrt{1-{\omega_p}^2/{\omega_0}^2}$, where $\omega_p$ and $\omega_0$ are the plasma and laser angular frequencies respectively.
For a linear relativistic plasma wave, (i.e.~with wavelength $\lambda_p = 2\pi c/\omega_p$ and Lorentz factor $\gamma_\phi \approx \sqrt{n_c/n_e}\gg 1$, where $n_c =\epsilon_0  m_e {\omega_0}^2 / e^2$ is the critical density), the dephasing length is $L_{\rm d} = \gamma_{\phi}{^2} \lambda_p = (\omega_0/\omega_p)^{2}\lambda_p$.

For a short duration laser pulse driving a nonlinear wakefield, (pulse length $\sigma_t < \lambda_p/c$ and normalized vector potential $a_0 \gg 1$), plasma electrons are pushed outward by the front of the pulse such that the rear of the pulse propagates in an ion cavity. 
Pump depletion occurs at the front of the laser pulse as energy is coupled into the plasma wave or lost due to diffraction. 
As a result of this localized depletion, the laser rapidly evolves to have a sharp rising edge, which etches back through the pulse \cite{Decker1996POP, Vieira2010NJP}. Decker {\it et al.} \cite{Decker1996POP} showed that the velocity of this pulse front etching is $v_{\rm etch} = c {\omega_p}^2/{\omega_0}^2$ in the group velocity frame of the laser. The depletion length is then
\begin{equation}
L_{\rm{dp}} \approx ({\omega_0}^2/{\omega_p}^2) \sigma_t c.
\label{eq:NLdepletionLength} 
\end{equation}
For a near-resonant pulse, $c\sigma_t\approx\lambda_p$, then $L_{\rm dp} \approx L_{\rm d}$, and so depletion should not limit electron energy gain. However, because of pulse front etching, the effective laser pulse velocity is reduced, such that the plasma wave phase velocity becomes $v_\phi/c = (v_g - v_{\rm etch})/c \approx 1-\frac{3}{2}(\omega_p{^2}/\omega_0{^2})$ for $\omega_p \ll \omega_0$. 
This reduces the dephasing length to $L_{\rm d} = \frac{2}{3}(\omega_0/\omega_p)^{2}\lambda_p$, which limits the maximum electron energy gain \cite{Lu2007PRSTAB}. 

Although guiding has been demonstrated up to $L_{\rm dp}$ for a range of plasma densities \cite{Ralph2009PRL}, no quantitative measurement of pump depletion in nonlinear LWFA has been reported. 
In addition, the pulse front etching model does not include laser power amplification that is observed in numerical simulations \cite{Gordon2003PRL}. Pulse shortening of LWFA drive pulses has been previously reported \cite{Faure2005PRL,Schreiber2010PRL}, but not the  power amplification, which is vital for self-injection of electrons \cite{Kalmykov2009PRL, Froula2009PRL, Kneip2009PRL, Savert2015PRL, Bloom2018PRL}.

In this letter, we present measurements of energy depletion and pulse compression of a relativistic ($a_0 \gg 1$) short-pulse ($c \sigma_t < \lambda_p$) laser in a self-guided \mbox{LWFA}. We also report the first direct measurements of power amplification of the driving laser pulse. These results are modeled by considering group velocity dispersion of the laser pulse as the leading edge is redshifted. The power amplification was found to be coincident with the onset of electron self-injection, confirming its vital role in this process.

The experiment 
was performed using the Gemini laser \cite{Hooker2006JDP}, interacting with a supersonic helium gas jet, at electron densities of up to \unit[$n_e=4\times 10^{18}$]{cm$^{-3}$} (\unit[0.007]{$n_c$}). 
Each linearly polarized laser pulse, of wavelength $\lambda_0=\unit[800]{nm}$, contained \unit[$11 \pm 1$]{J} of energy in a duration of \unit[$t_{\rm FWHM}=51 \pm 3$]{fs}. The laser was focused onto the front of the gas target with an $f/20$ parabolic mirror at a peak $a_0 \simeq 3.0$.

A magnetic electron spectrometer was used to measure the spectrum of the accelerated electron beam. The laser pulse at the exit plane of the plasma was imaged with a pair of $f/10$ spherical mirrors at a resolution of \unit[10]{$\mu$m} over a field of view of \unit[902]{$\mu$m} $\times$ \unit[675]{$\mu$m}. The transmitted energy was measured by integrating the counts on the camera, which was cross-calibrated with an energy diode. Also, a \unit[5]{mm} diameter area near the center of the transmitted beam, $\approx 1/20^{\rm th}$ of the full beam diameter, was directed to two Grenouille (Swamp Optics) SHG-FROGs (second harmonic generation frequency resolved optical gating) \cite{Trebino2000Book}. These devices produce spectrally dispersed autocorrelations, from which the complete temporal intensity and phase information of the pulse was retrieved using an iterative algorithm. 

\begin{figure}[!t] 
   \centering
   \includegraphics[width=8.5cm]{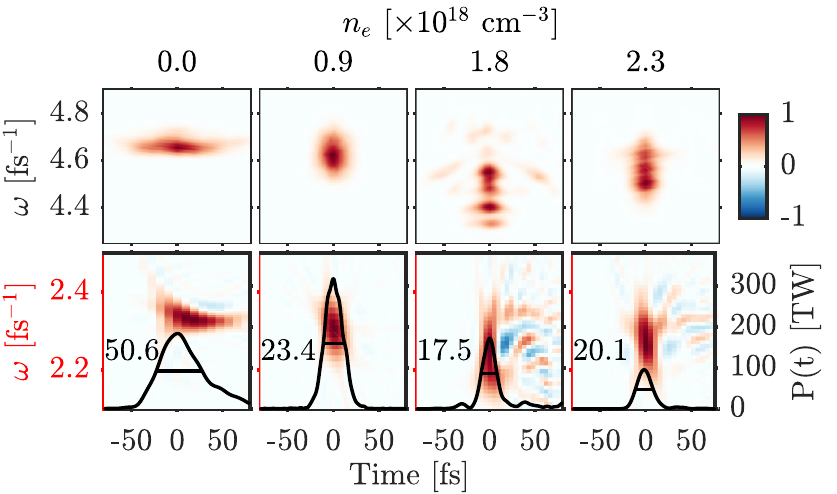} 
   \caption{(Top) FROG traces and (bottom) Wigner transforms with temporal profiles (black lines) for (left-to-right) \unit[$n_e = (0, 0.9, 1.8, 2.3) \times 10^{18}$]{cm$^{-3}$} and a nozzle diameter of \unit[15]{mm}. The Wigner transforms and temporal profiles have been corrected for diagnostic dispersion, representing the pulse at the plasma exit. The fs FWHM pulse durations are displayed by each pulse.}
   \label{fig:PowerAmp_frogWigs}
\end{figure}
The spectral window of the FROG diagnostics were limited, such that the first could measure to a lower pulse length limit of \unit[10]{fs}, for a time-bandwidth limited pulse, while the second was restricted to \unit[20]{fs}. 
Additional glass was placed in the beam path of the second FROG to create a known spectral phase offset between the two diagnostics. 
Only retrieved pulses with the correct time direction have the correct phase offset, and so the inherent time-direction ambiguity could be resolved.
This process was possible for \unit[$n_e < 0.6 \times 10^{18}$]{cm$^{-3}$}, while at higher densities, the spectrum became too broad for the second FROG. 
For these measurements, gradual changes of the pulse shape and Wigner \cite{Wigner1932PR} transforms with increasing density were used to determine the direction of time.
The phase retrieval algorithm was performed 10 times, each time with a different random seed, and variations in the retrieved pulses were included in the measurement error. Shots with visibly poor retrievals, large FROG errors (rms relative pixel error $>0.02$) or unresolved time direction uncertainties are not included in the results. Out of 59 shots, 43 are included in the graph. 
%

Example FROG traces and retrieved pulses are shown in \cref{fig:PowerAmp_frogWigs}. The pulses were observed to frequency downshift and temporally compress for increasing plasma density up to \unit[$n_e = 1.5 \times 10^{18}$]{cm$^{-3}$} for a \unit[15]{mm} diameter nozzle. Beyond this density, the pulse length increased again due to energy depletion of the laser pulse.

\begin{figure}[!thp] 
   \centering
   \includegraphics[width=8.5cm]{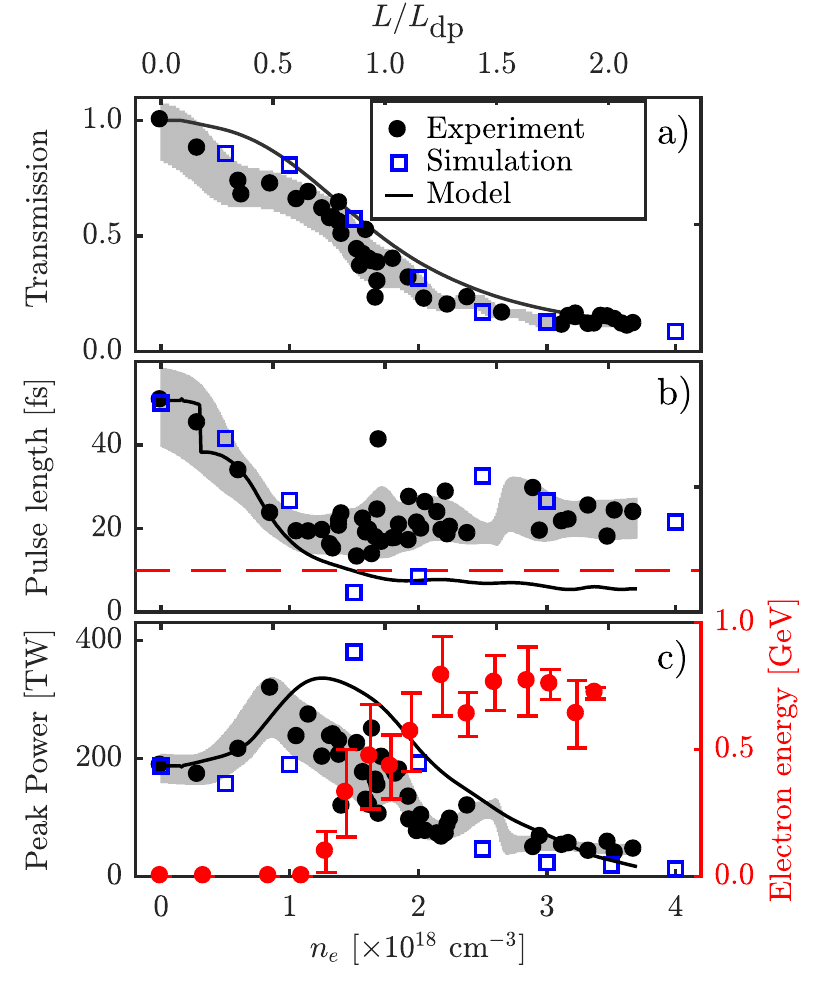} 
   \caption[Pulse front etching model comparison to measure pulse depletion]{Experimental and simulated \textbf{a} transmitted laser energy fraction, \textbf{b} pulse duration (FWHM) and \textbf{c} peak pulse power and maximum observed electron beam energy (red circles), versus plasma density for a \unit[15]{mm} nozzle diameter. The gray shaded regions indicate rms error (statistical and measurement errors) of a moving average of the data points. The red dashed line in \textbf{b} is the instrument limit of the FROG for time-bandwidth limited pulses. The solid black lines are calculated from our pulse evolution model.}
   \label{fig:PowerAmp_ETP}
\end{figure}

The results are shown as a function of plasma density for a \unit[15]{mm} nozzle diameter in \cref{fig:PowerAmp_ETP}. For the plots of pulse energy transmission, transmitted pulse length and peak power, each data point represents one measurement. The maximum electron energy [\cref{fig:PowerAmp_ETP}c] was taken as the highest point where the electron signal was above three times the rms background variation, after subtraction of the on-shot background. These values are averaged over multiple ($\sim 4$) shots within a density bin width of \unit[$0.2 \times 10^{18}$]{cm$^{-3}$}, with error bars combining statistical and measurement errors.
The ratio of the interaction length to $L_{\rm dp}$ [\cref{eq:NLdepletionLength}] is shown for comparison at the top of \cref{fig:PowerAmp_ETP}a, with $\sigma_t = t_{\rm FWHM}/\sqrt{2 \ln(2)}$.

Transmitted laser energy decreased with increasing density [\cref{fig:PowerAmp_ETP}a]. About 50\% of the energy was transmitted for $L=L_{\rm dp}$, and was only fully depleted for $L \simeq 2L_{\rm dp}$. Imaging of the exit plane showed a guided spot for \unit[$n_e > 1 \times 10^{18}$]{cm$^{-3}$}, while for \unit[$n_e > 3 \times 10^{18}$]{cm$^{-3}$}, only the unguided fraction of the laser energy was observed. 
The measured pulse duration [\cref{fig:PowerAmp_ETP}b] decreased from an initial \unit[$51 \pm 3$]{fs} to \unit[$\sim 20$]{fs} for $n_e >10^{18}$. 
For these higher densities, the spectral broadening resulted in spectral clipping in the FROG diagnostic and so the retrieved pulse was longer than the input pulse.
The shortest observed pulse length in a single shot was \unit[$13.0 \pm 1.3$]{fs}  for \unit[$n_e = 1.5 \times 10^{18}$]{cm$^{-3}$}. 
The peak power of the laser pulse after the interaction [\cref{fig:PowerAmp_ETP}c] was calculated by setting the energy of the transmitted pulse equal to the time integral of the temporal pulse shape. The power was observed to increase, with a maximum at \unit[$n_e = 0.9 \times 10^{18}$]{cm$^{-3}$}, increasing from \unit[$187 \pm 11$]{TW} to \unit[$318 \pm 12$]{TW}. 
Wide-angle electron emission was produced for \unit[$n_e > 0.2 \times 10^{18}$]{cm$^{-3}$} but the charge increased significantly for \unit[$n_e > 1.1 \times 10^{18}$]{cm$^{-3}$} [\cref{fig:PowerAmp_ETP}c], where the maximum power enhancement was observed. Maximum electron energies, of \unit[$0.79 \pm 0.12$]{GeV}, occurred at \unit[$n_e \simeq 2.3 \times 10^{18}$]{cm$^{-3}$}, in line with scaling predictions \cite{Lu2007PRSTAB}. 

\begin{figure}[!htp] 
\includegraphics[width=8.5cm]{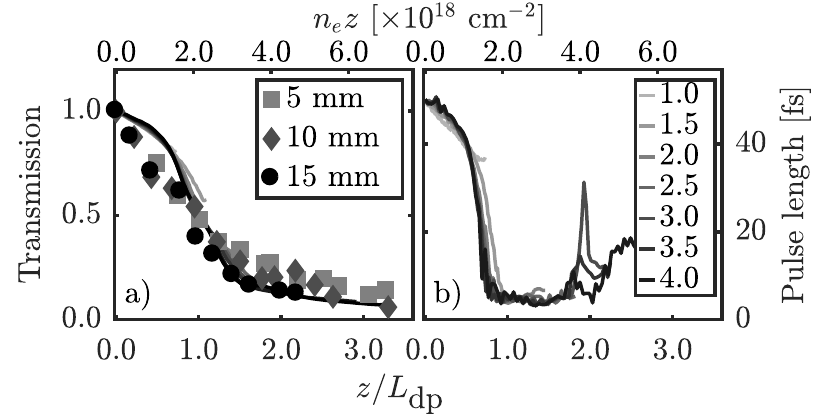} 
\caption[Laser evolution scales with areal density]{Laser depletion \textbf{a} and pulse compression \textbf{b} as functions of the areal density (top axis), and the interaction length normalized by the depletion length (bottom axis). Experimental measurements (markers) are averaged over a \unit[$0.25 \times 10^{18}$]{cm$^{-2}$} bin width for three different nozzle diameters (inset \textbf{a}). Simulated values (lines) are given for different electron densities (inset \textbf{b} in units of \unit[$10^{18}$]{cm$^{-3}$})}
   \label{fig:PowerAmp_arealDensity}
\end{figure}

\Cref{fig:PowerAmp_arealDensity} shows laser measurements from three nozzle diameters (temporal diagnostic was only available for the \unit[15]{mm} nozzle). 
The similarity of the results when scaling the propagation length by the depletion length [\cref{eq:NLdepletionLength}] indicates that laser evolution is a function of areal density, $z n_e \propto z/L_{\rm{dp}}$, over the covered density range.

The experiment was simulated using the OSIRIS \cite{Fonseca2002ICCS} particle-in-cell code in 2D3V geometry. The simulation window moved at $c$ along the laser propagation direction and had dimensions of \unit[$200 \times 200$]{$\mu$m} divided into \unit[$8000 \times 800$]{cells} in the pulse propagation ($z$) and the transverse ($x$) directions respectively.
The pulse envelope was modeled using a polynomial approximation to a gaussian with \unit[$\tau_{\rm FWHM} =50$]{fs}, focused to a spot width \unit[$w_{\rm FWHM}=25$]{$\mu$m} and a peak $a_0$ of 3.0. The plasma target was \unit[15]{mm} in length, including linear density ramps over \unit[$500$]{$\mu$m} at the entrance and exit of the plasma, approximating the experimental density profile, with 4 electron macro-particles per cell and stationary ions. 

The simulated pulse properties at the plasma exit are shown alongside the experimental data in \cref{fig:PowerAmp_ETP,fig:PowerAmp_arealDensity}. The energy depletion and pulse compression proceeds at similar rates as in the experiment, but the pulse compresses to a minimum of \unit[4]{fs} (below the experimental measurement limit) at \unit[$n_e = 1.5 \times 10^{18}$]{cm$^{-3}$}. At higher densities, the pulse length increases again, once almost all the laser energy depletes, and the short compressed peak in the laser field vanishes.

\begin{figure}[!htp] 
\includegraphics[width=8.5cm]{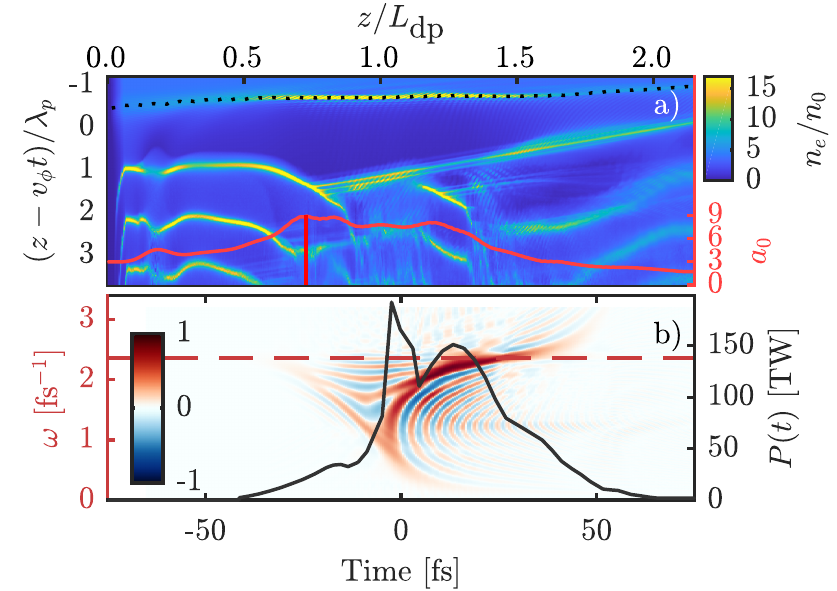} 
\caption[Laser and plasma evolution in a PIC simulation]{\textbf{a} On-axis electron density map (image) in a frame moving at $v_{\phi} = v_g - v_{\rm etch}$ and peak $a_0$ of the laser (red line), as functions of propagation distance in a simulation with \unit[$n_e = 4 \times 10^{18}$]{cm$^{-3}$}. The first maxima of the plasma wave, is overlaid with a black dashed line. \textbf{b} Wigner transform (blue-red) of the laser pulse overlaid with the temporal intensity profile at $z =0.7 L_{\rm dp}$ (\unit[3.9]{mm}). The red horizontal line shows the initial laser frequency.}
   \label{fig:simWakeEvol}
\end{figure}

\Cref{fig:simWakeEvol}a shows the propagation dependence of the on-axis plasma density modulation in the reduced group velocity ($v_{\phi} = v_g - v_{\rm etch}$) reference frame, for \unit[$n_e = 4 \times 10^{18}$]{cm$^{-3}$}. The laser peak $a_0$, plotted as a red line, first increases via pulse compression, and then decreases due to pump depletion. Self-focusing is observed in the first \unit[1]{mm} of plasma, after which a stable guided spot size is reached. 
The density peak coinciding with the leading edge of the laser pulse moves at close to the reduced group velocity, as represented by a horizontal line in the coordinate frame of the figure, in agreement with the pulse front etching model \cite{Decker1996POP}. 
Deviation from this velocity is seen at early times, before the sharp front of the driving pulse forms, and at late times, once the laser is mostly depleted. 

Rapid compression of the driving pulse occurs \unit[4]{mm} into the target, increasing the plasma bubble radius, since $r_b \propto \sqrt{a_0}$ \cite{Lu2006PRL}. During this stage of the interaction, the effective phase velocity of the back of the wake decreases from $\gamma_{\phi}=20$ to $\gamma_{\phi}=7$. This coincides with self-injection of plasma electrons \cite{Schroeder2011PRL}, as seen by the straight lines originating from the back of the first plasma wave and advancing relative to the plasma wave as the simulation progresses. The injection occurs over a short propagation distance, populating a narrow phase region of the wakefield, similar to injection mechanisms that use tailoring of the target density profile to modify the plasma wave phase velocity \cite{Bulanov1998PRE, Geddes2008PRL,Kalmykov2009PRL}.

The pulse frequency-shift and compression is illustrated by the Wigner transform of the simulated laser pulse at \unit[$z=4$]{mm} in \cref{fig:simWakeEvol}b. The pulse is largely red-shifted at the first density maximum of the plasma wave, close to the maximum of the laser intensity, as discussed by Schreiber \textit{et al.} \cite{Schreiber2010PRL}. 


In the simple picture of pulse front etching, the leading edge of the laser pulse continually moves back through the pulse, locally reducing the laser power to zero. 
In reality, as photons at the leading edge are redshifted they will begin to slip back through the pulse due to group velocity dispersion, as illustrated in \cref{fig:simWakeEvol}b. Photons are able to drift back away from the depletion region once their velocity is less than that of the laser pulse front. Equating the group velocity of a redshifted photon to the reduced group velocity of the laser pulse front, $v_g(\omega_{\rm min}) = v_g(\omega_0) - v_{\rm etch}$, gives the minimum frequency reached by these photons as $\omega_{\rm min} = \omega_0/\sqrt{3}$. Once the power of the leading edge of the laser pulse drops below the critical power for self-focusing $P_c$, then it will diffract and no longer drive a plasma wave. Therefore, the energy coupled into the plasma from a local region in the power profile $W = (1-1/\sqrt{3})(P(t) -P_c) \Delta{t}$. The remaining energy, $(1/\sqrt{3})(P(t) -P_c) \Delta{t}$, is transported back through the pulse by these redshifted photons, into a region of low plasma density, where the group velocity dispersion is much smaller. This leads to an increase in power behind the depletion front, and thereby modifies the energy depletion rate. 

Numerical calculations were performed, by stepping though the initial power profile $P(t)$ from the first point at which $P(t) > P_c$ for a given plasma density, reducing the power at this point to $P(t) = P_c$. The un-depleted fraction of this energy, ($(1/\sqrt{3})(P(t) -P_c) \Delta{t}$), is added to the following region of the pulse, averaged over $\lambda_p/4$ to approximate the effect of group velocity dispersion. The pulse energy, duration and peak power after propagating \unit[13]{mm} are calculated for each plasma density and are plotted as black lines in \cref{fig:PowerAmp_ETP}.

The numerical model predicts the energy depletion rate observed experimentally. The pulse compression is well reproduced until \unit[$n_e > 1 \times 10^{18}$]{cm$^{-3}$}, where the pulse length reaches a value lower than can be measured experimentally. 
The power amplification effect is also matched by the model, predicting a maximum power for \unit[$n_e = 1.3 \times 10^{18}$]{cm$^{-3}$}. 
Though, at this density, the experimental measurement is instrument limited, because the pulse spectrum was broader than spectral range of the FROG diagnostic. 

For a gaussian laser pulse with the initial peak power $P_0 > P_c$, maximum power amplification is reached when the pulse is etched to approximately the midway point, and the laser pulse energy is reduced by $\sim 50\%$. This occurs after an evolution length,
\begin{align}
L_{\rm evol} = \sigma_t c \left(\frac{2}{3}\frac{{\omega_0}^2}{{\omega_p}^2}\right)\sqrt{\frac{1}{2}\ln\left(\frac{P_0}{P_c}\right)} \;,
\end{align}

For moderate values of $P_0/P_c$ (\unit[$n_e>0.6 \times 10^{18}$]{cm$^{-3}$} in \cref{fig:PowerAmp_ETP}), the evolution length is approximately equal to the usually quoted depletion length $L_{\rm dp}$, and the pulse is only fully depleted at $L \approx 2 L_{\rm evol}$. 
Taking injection to occur first at the position $z_{\rm inj}=L_{\rm evol}$, pump depletion of a \unit[180]{TW} \unit[50]{fs} pulse occurs before dephasing for \unit[$n_e<4.2 \times 10^{18}$]{cm$^{-3}$}. 
At the density for which the maximum electron energy was observed in the experiment, \unit[$n_e = 2.3 \times 10^{18}$]{cm$^{-3}$}, injection occurs at \unit[$z_{\rm inj}=8.6$]{mm}, giving an acceleration length of \unit[4.4]{mm}. 
However, the initial pulse shape in the experiment was non-gaussian, having a rapid rising edge and an extended falling edge. 
As a result, $L_{\rm evol}$ was shortened to \unit[7.2]{mm}, with the consequence that more laser energy remained in the pulse, allowing the acceleration length to be extended to \unit[5.8]{mm}. 
Using this value, and the experimentally measured electron energy, the average acceleration gradient over this acceleration length was \unit[$\approx 140$]{GeVm$^{-1}$}.

Hence, our pulse evolution model demonstrates that the laser pulse evolution is heavily influenced by the initial pulse shape. By using a pulse with an initially sharp rising edge, the laser pulse peak power increases rapidly, triggering injection before much of the laser energy is lost. A slow falling edge will then extend the depletion length, which could allow a large $a_0$ to be maintained over a longer distance. In this way, it may be possible to tailor the pulse shape to optimize the injection and acceleration processes, benefiting the many applications of these accelerators, such as the generation of large numbers of x-rays \cite{Kneip2011NP}, gamma rays \cite{Sarri2015NC} and positrons \cite{Sarri2014PRL}.

We would like to acknowledge technical support from CLF staff.
This research was supported by STFC (ST/P002056/1, ST/J002062/1, ST/P000835/1), and  EPSRC (EP/I014462/1). 
We thank the \textsc{Osiris} consortium (UCLA/IST) for the use of \textsc{Osiris}. M. Wing acknowledges the support of DESY, Hamburg and the Alexander von Humboldt Stiftung.
A.G.R. Thomas acknowledges funding from the NSF grant 1535628 and DOE grant DE-SC0016804.
All data created during this research are openly available from Lancaster University data archive at http://dx.doi.org/10.17635/lancaster/researchdata/15.
This work has now been published at 

$^*$ z.najmudin@imperial.ac.uk

\end{document}